\documentclass[12pt]{JHEP} 

\usepackage{epsfig}
\usepackage{bbold}

\newcommand{\Tr}{{\rm Tr}\ }
\newcommand{\uni}{\mathbb{1}}

\newcommand{\W}{{\cal W}}

\newcommand{\Path}{\mbox{P}}

\newcommand{\same}[1]{\hspace{.2cm}\raisebox{-0.4cm}
{\epsfxsize=1cm\epsfbox{#1.eps}}
\hspace{-1.2cm}\raisebox{-0.3cm}{\epsfxsize=.35cm\epsfbox{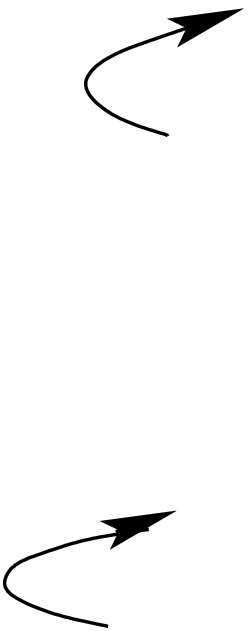}}
\hspace{1.2cm}}

\newcommand{\feynline}[1]{\raisebox{-0.2cm}{\epsfxsize=0.7cm\epsfbox{#1.eps}}}
\newcommand{\be}{\begin{equation}}
\newcommand{\ee}{\end{equation}}
\newcommand{\bea}{\begin{eqnarray}}
\newcommand{\eea}{\end{eqnarray}}

\newcommand\nn{\nonumber}

\newcommand{\eqn}[1]{(\ref{#1})}
\newcommand{\xd}[1]{\dot x_{#1}}

\title{Limiting Geometries of Two Circular Maldacena-Wilson Loop
Operators} 

\author{Gleb Arutyunov, Jan Plefka and Matthias Staudacher \\
Albert-Einstein-Institut, Max-Planck-Institut
f\"ur Gravitationsphysik\\Am M\"uhlenberg 1, D-14476 Golm, Germany\\
Email: \email{agleb,plefka,matthias@aei-potsdam.mpg.de}}

\preprint{\hepth{0111290}\\
AEI-2001-140\\}

\abstract{
We further analyze a recent perturbative two-loop calculation 
of the expectation value of two axi-symmetric circular 
Maldacena-Wilson loops in ${\cal N}=4$ gauge theory. Firstly, it is 
demonstrated how to adapt the previous calculation of 
anti-symmetrically oriented circles to the symmetric case.
By shrinking one of the circles to zero size we then explicitly 
work out the first few terms of the local operator expansion
of the loop.  Our calculations explicitly demonstrate that 
circular Maldacena-Wilson loops are non-BPS observables precisely
due to the appearance of unprotected local operators. The latter
receive anomalous scaling dimensions from non-ladder diagrams.
Finally, we present new insights into a recent conjecture  
claiming that coincident circular Maldacena-Wilson loops are 
described by a Gaussian matrix model. We report on a novel, 
supporting two-loop test, but also explain and illustrate why 
the existing arguments in favor of the conjecture are flawed.
}


\begin{document}

\section{Introduction and Conclusions}

Wilson loops are interesting non-local observables in gauge theories.
They are functionals of macroscopic space-time contours and are 
widely suspected to be the proper observables to describe 
the strong coupling physics of Yang-Mills theories. While they have been
quite useful as order parameters in lattice gauge theory, Wilson loops
have unfortunately been rather difficult to work with in continuum gauge
theories. One of the chief reasons is that they are plagued by
various infinities whose removal obscures the seeming simplicity of their
bare equations of motion. The form of the renormalized,
non-perturbative loop equations remains unknown.

For smooth loop contours the mentioned infinities can be divided into 
two classes: In addition to the usual quantum field theoretic ``bulk''
divergences stemming from Feynman diagrams containing internal loops
there are further ``boundary'' divergences due to contact interactions
on the boundaries of the Wilson loops. 

Recently, a much more transparent situation has begun to emerge in
maximally supersymmetric gauge theory. In \cite{Maldacena:1998im}
a modification of the usual Wilson loop operators has been
proposed. Here the loops are not only coupled to the gauge
field but in addition to the six scalar fields of the model. 
This modification significantly improves the just mentioned
divergence problems: The boundary divergences of individual
Feynman graphs are either absent or, even more interestingly, cancel
against bulk divergences. This phenomenon has so far been observed
in a number of one-loop and two-loop calculations 
\cite{Erickson:2000af}, \cite{Plefka:2001bu}.
However, a general proof of the perturbative, let alone non-perturbative,
finiteness of these novel Maldacena-Wilson loop operators is still
lacking to date. Finding such a proof would yield an infinite
set of finite geometric probes of a four-dimensional quantum field
theory.

A further - actually the initial - motivation for being interested 
in these operators is that they are conjectured to be directly
related, at strong coupling and in the so-called planar limit, to
certain classical supergravity solutions in a special background
\cite{Maldacena:1998im}. Finally, there is
even hope that for some special contours such as a circle these
loops might be exactly calculable \cite{Erickson:2000af,
Drukker:2000rr}, which, if proven, might lead 
to non-trivial analytic tests of the proposed
supergravity-gauge correspondence for these macroscopic observables. 
To date the correspondence 
\cite{Maldacena:1998re,Gubser:1998bc,Witten:1998qj} 
has only been rigorously tested on the level of certain 
local operators. 

In \cite{Plefka:2001bu} the first two-loop perturbative study of
Maldacena-Wilson loops was reported. The specific situation
analyzed consisted in two axi-symmetric circular contours of
arbitrary radii and distance. The motivation of \cite{Plefka:2001bu}
was to perform a two-loop test of the finiteness properties, as well
as to establish that the static potential, obtained by sending
the radii to infinity, receives contributions
from non-ladder diagrams, in contradistinction to an accidental
one-loop cancellation of interactive diagrams.
For related work on Maldacena-Wilson loops see
\cite{Drukker:1999zq}-\cite{BGK}.

In the present paper the results of \cite{Plefka:2001bu} are applied
to two further situations of physical interest. This requires
a rather straightforward extension of our previous results,
which were derived for circles of opposite orientation, to also
include the equal-orientation case (see section 2).
We find that it simply corresponds to formally flipping the relative
sign of the two radii.
In section 3 we then study, for both orientations, the limit where 
either one of the two circles shrinks to zero size. In this limit one
expects that the shrinking loop should be expandable in local
operators of increasing scaling dimension. This allows to extract
term-by-term the expectation value of local operators with the
remaining ``large'' circle. It is shown that our previous results
contain enough information to extract the one-loop anomalous dimension
of the fields of approximate scaling dimension two and three. 
In particular we recover the known one-loop anomalous dimension
of the Konishi field, which is the lowest ($\sim 2$) dimension 
unprotected operator of the theory. More importantly, it is reassuring
that the anomalous scaling turns out to be precisely due to the
{\it non-ladder diagrams}. In turn, it is seen that
the unprotected operators appearing in the
operator expansion are responsible for the global non-BPS nature of closed
Maldacena-Wilson contours. We also comment on the appearance of
unprotected operators of classical dimension three and four and
their correspondence to the supergravity limit. 

The second application, presented in section 4,
involves the limit of equal but finite
radii and vanishing distance of the equally oriented circles\footnote{
This is the identical to the ``static potential'' limit considered in
\cite{Plefka:2001bu} except that there the orientation was 
anti-symmetric.}.
This allows us to perform the first two-loop test of a conjecture
due to Drukker and Gross \cite{Drukker:2000rr} (see also 
\cite{Erickson:2000af}), which holds that multiple coincident circular
Maldacena-Wilson loops can be described by a Gaussian matrix model.
We find that all connected non-ladder diagrams cancel to 
${\cal O}(g^6)$ for the case of two circles, in line with the
conjecture.

We nevertheless feel the need to add some cautionary observations.
The fact that so far the Gaussian matrix model description appears to
be valid might be a purely accidental low-order phenomenon. At any
rate we argue that no real arguments exist that vertex diagrams should
not contribute to the circle, and that the analysis presented in
\cite{Drukker:2000rr}, which involves a conformal map of an 
infinite line to a circle, does not directly apply to vertex graphs.
As an illustration we consider a
particular two-loop non-ladder graph for a single circle and demonstrate
that it is, just like the ladder graphs, zero for the line but
perfectly finite and non-zero for the circle. It might still
be true that the {\it sum} of all non-ladder diagrams cancels, but
we so far lack any argument for this. Short of a general proof,
it would clearly be desirable to complete the full
two-loop calculation for a single circle.

\section{Two Loops of Equal Orientation}

In \cite{Plefka:2001bu} a two loop perturbative
calculation of the (connected) expectation value of two circular
Maldacena-Wilson loops of {\it opposite} orientation
was performed. Here we shall be interested in the
scenario with two loops of {\it equal} orientation.
That is we consider the connected correlator
of two Maldacena-Wilson loops
\be
\label{2loopcor}
\langle \W(C_1)\, \W(C_2) \rangle_{c}
= \langle \W(C_1)\, \W(C_2) \rangle-
\langle \W(C_1)\rangle \, \langle\W(C_2) \rangle \, .
\ee
The Maldacena-Wilson loop
operator is defined by
\be
\W[C]=\Tr \Path \exp \Bigl [\oint_C d\tau(iA_\mu(x)\dot x^\mu + \Phi_I(x)\theta^I |\dot x|)
\Bigr ]
\label{maldaloop}
\ee
Here $\theta^I$ is a point on the unit five-sphere,
i.e.~$\theta^I\theta^I=1$, and
$x^\mu(\tau)$ parameterizes the curve $C$.
We take the curves $C_1$ and $C_2$ to be two parallel, axi-symmetric
circles of equal
orientation and, respectively,
radii $R_1$ and $R_2$ separated by a distance $h$
\bea
x^\mu(\tau)&=& ( R_1\, \cos \tau, R_1\, \sin\tau,\, 0\, ,\,0)  \nn\\
y^\mu(\sigma)&=& ( R_2\, \cos \sigma, R_2\, \sin\sigma, \, h \, ,\,0)
\qquad \tau,\sigma\in [0,2\pi]
\label{parametrization}
\eea
All our conventions follow  \cite{Plefka:2001bu}.
For calculational purposes it is
useful to go ten dimensional notation $M=(\mu,I)$,
${\cal A}_M^a[x]=(A_\mu^a(x),\Phi_I^a(x))$
and ${\dot x}(\tau)=({\dot x}^\mu(\tau),\theta^I\, |{\dot x}^\mu|)$
with
the combined gluon-scalar propagator in $2\omega$
dimensions
\be
\langle {\cal A}_M^a[x]\, {\cal A}_N^b[y] \rangle = g^2\,
\delta^{ab}\, \delta_{MN}
\frac{\Gamma(\omega-1)}{4\pi^\omega}\,
\frac{1}{[(x-y)^2]^{\omega-1}}
\ee
in Feynman gauge where $x$ and $y$ are points in $2\omega$
dimensional Euclidean space. As a matter of fact
the results of \cite{Plefka:2001bu} may be directly
translated to our new scenario of circles of equal
orientation\footnote{We wish to thank N.~Drukker for
independently suggesting that our calculation could 
be easily modified to also cover
this situation. He also, correctly, proposed that the 
equal orientation case could be obtained by
simply flipping the sign of one of the two radii.}.
To do so consider the integrated
Maldacena-Wilson loop associated with $R_2$
with an open leg at the point $x^\mu$
\bea
\raisebox{-.5cm}{\epsfxsize=1.8cm\epsfbox{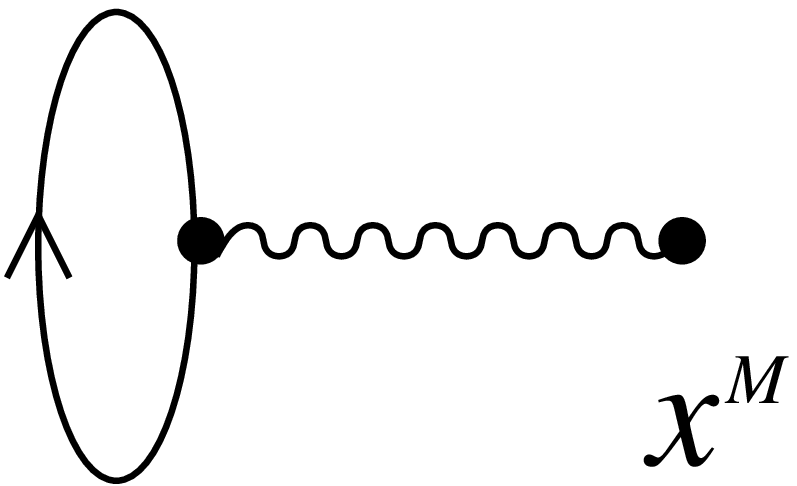}}
&=& \int_0^{2\pi} d\sigma \,
{\dot y}(\sigma)\,
\langle \,
{\cal A}_M^a[y(\sigma)]\, {\cal A}_N^b[x] \,\rangle \nn\\
&=&\frac{g^2\,\delta^{ab}\, \Gamma(\omega-1)}{4\pi^\omega}\,\int_0^{2\pi}d\sigma\,
\frac{({\dot y}^\mu(\sigma),\theta^I R_2)}{[(y(\sigma)-x)^2]^{\omega-1}} \nn\\
&=& \frac{g^2\,\delta^{ab}\, \Gamma(\omega-1)}{4\pi^\omega}\,\int_0^{2\pi}d\sigma\,
\frac{R_2\, ( - \cos \sigma\, \sin \phi , \cos\sigma\, \cos \phi, \, 0\, ,\,  0\,
 , \theta^I)}
{(A^2 -2R_2\, \rho\, \cos \sigma)^{\omega-1}}
\label{circle-a}
\eea
where we have introduced polar coordinates
$x_1=\rho\, \cos \phi$, $x_2=\rho\, \sin \phi$ and $A^2=R_2^2+ \rho^2
+(x_3-h)^2+ x_4^2$. Moreover a shift in $\sigma\rightarrow \sigma +\phi$
was performed. Now consider the same graph with a circle of
opposite orientation, parametrized by
\be
\bar y^\mu(\sigma)= ( R_2\, \cos \sigma, - R_2\, \sin\sigma, \, h \, ,\,0)  \, .
\ee
Performing the analogous manipulations as in \eqn{circle-a} one arrives
at
\bea
\raisebox{-.5cm}{\epsfxsize=1.8cm\epsfbox{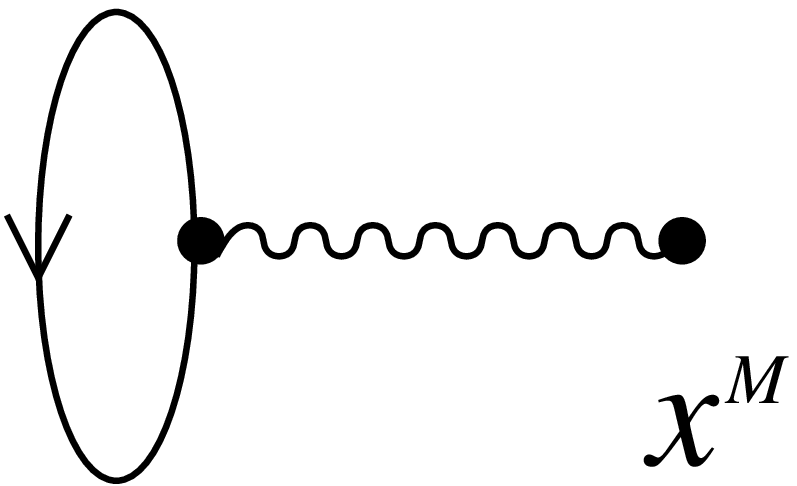}}
&=&\frac{g^2\,\delta^{ab}\, \Gamma(\omega-1)}{4\pi^\omega}\,\int_0^{2\pi}d\sigma\,
\frac{({\dot {\bar y}}^\mu(\sigma),\theta^I R_2)}{[({\bar y}
(\sigma)-x)^2]^{\omega-1}} \nn\\
&=& \frac{g^2\,\delta^{ab}\, \Gamma(\omega-1)}{4\pi^\omega}\,\int_0^{2\pi}d\sigma\,
\frac{R_2\, ( \cos \sigma\, \sin \phi , -\cos\sigma\, \cos \phi, \, 0\, ,\,  0\,
 , \theta^I)}
{(A^2 -2R_2\, \rho\, \cos \sigma)^{\omega-1}}
\label{circle-b}
\eea
Now clearly \eqn{circle-b} may be transformed into \eqn{circle-a} by
first shifting the integration variable $\sigma\rightarrow \sigma +\pi$,
thereafter sending $R_2\rightarrow -R_2$ and multiplying the
total graph by $-1$. We have hence shown that
\be
\raisebox{-.5cm}{\epsfxsize=1.8cm\epsfbox{1CircleBar.eps}} = -\, \Bigg[\,
\raisebox{-.5cm}{\epsfxsize=1.8cm\epsfbox{1Circle.eps}}
\,\Bigg ]_{R_2\rightarrow -R_2}
\label{Rrule}
\ee
This reasoning is applicable to all graphs where there is 
at least one circle
with no occurrence of path ordering along the loop. By looking at
the relevant graphs up to order $g^6$ in \cite{Plefka:2001bu}
we see that the rule \eqn{Rrule} is applicable to all graphs except for
the 3-ladder diagrams $\feynline{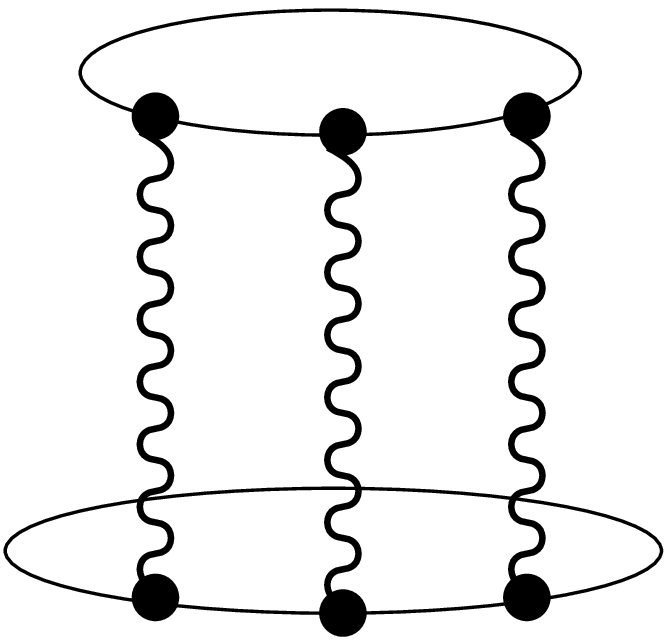}$ and $\feynline{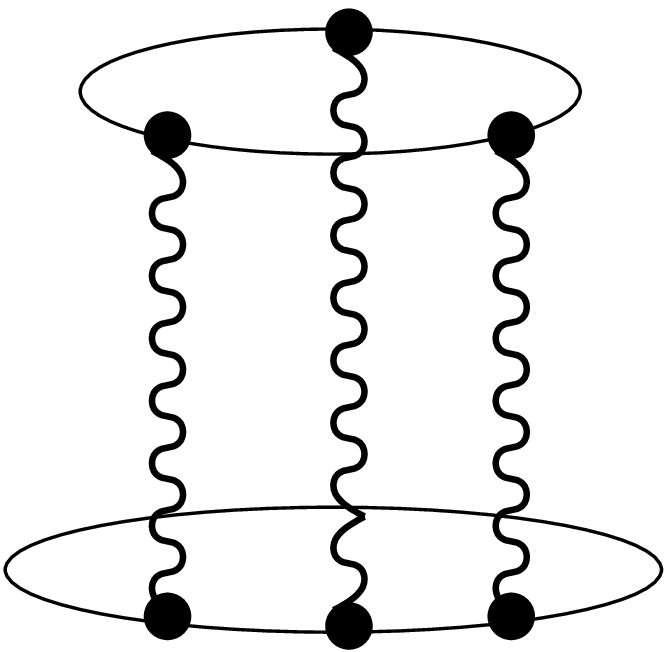}$.
A careful reanalysis of these two diagrams, however, reveals that also
for these two path ordered diagrams the rule \eqn{Rrule} continues
to hold.

In fact one wonders whether the rule \eqn{Rrule} holds in general,
i.e.~for any graph. However, we have not yet found a proof due to
the path ordering problems.

In summary we now have the complete two loop (${\cal O}(g^6)$)
computation for two connected Maldacena-Wilson loops of equal orientation
at our disposal. For this geometry we again observe the complete
finiteness of the Maldacena-Wilson loop, as the subtle
bulk-boundary cancellation of divergences of the self-energy
and the IY graphs of \cite{Plefka:2001bu} continues to hold
due to the simple rule \eqn{Rrule}.

\section{Local Operator Expansion}

We now turn to the study of our geometry when one of the two circles
shrinks to zero size, i.e.~the limit
$R_1\rightarrow 0$ while keeping  $R_2$ and $h$ finite. 
In \cite{Plefka:2001bu} we derived explicit analytic expressions for
all ladder graphs but had to content ourselves with somewhat
complicated if finite integral representations for the non-ladder diagrams.
Remarkably, in the present limit our expressions simplify sufficiently
and we are able to deduce explicit analytic results for all graphs.
Quite generally, one expects that the shrinking loop
may be represented by an infinite sum of local 
composite operators of increasing scaling dimension. 
The two loop results of \cite{Plefka:2001bu} allow to extract
informations on the leading lower dimensional 
operators appearing in this expansion and make contact 
with recent studies of local composite operators in ${\cal N}=4$
gauge theory performed in the context of the AdS/CFT correspondence
(see e.g. \cite{HFS}-\cite{PS}).

Consider the following local operator expansion 
of a Maldacena-Wilson loop
\bea \label{OPE}
\frac{W(C)}{\langle W \rangle }=\uni+\sum_k c_k R^{\Delta_k}O^{(k)}(x)
\, , 
\eea 
where $O^{(k)}$ denotes a local composite operator 
of scaling dimension $\Delta_k$. 
Assume the special case where the loop is a 
circle with radius $R$ and lies in 
the $(x_1,x_2)$ plane centered at $x$. Then the loop is sufficiently 
symmetric to only partially break Lorentz invariance. Thus 
the operator content in (\ref{OPE}) is classified by 
irreducible representations of $SO(2)\times SO(2)$,
the unbroken subgroup of $SO(4)$. Then every 
four-dimensional index $\mu$ naturally splits into 
$i$ and $\bar{i}$, where $i=1,2$
and $\bar{i}=3,4$. Moreover the loop possesses a definite orientation
captured by the  two-form $\epsilon^{ij}$ with $\epsilon^{12}=1$.

In the case where 
$R$ is small compared to any other distance in the problem at hand
the Maldacena-Wilson loop operator may be replaced by its local 
operator expansion. In particular, for the 
connected Green function 
of two parallel Wilson loops $C_1$ and $C_2$ with radii $R_1$ and $R_2$,
separated along the 3-axis by a distance $h$,
eq.(\ref{OPE}) implies the following decomposition 
\bea 
\label{2pt}
\frac{\langle W(C_1)W(C_2)\rangle_c} {\langle W \rangle^2
}&=&
\sum_k c_k R_1^{\Delta_k}\frac{\langle O^{(k)}(0)W(C_2)\rangle }{\langle W \rangle^2} \, 
\eea
assuming that $R_1\ll R_2,h$.
In perturbation theory the scaling dimension of an operator can
be represented as \bea \Delta=\Delta^{(0)} +
\Delta^{(1)}+\Delta^{(2)}+\ldots \, , \eea where $\Delta^{(0)}$ is
the free field dimension and $\Delta^{(1)}$, $\Delta^{(2)}$ are
anomalous dimensions at order $g^2$, $g^4$ and so on.
Therefore every term $R^{\Delta}$ in eq.(\ref{2pt})
produces logarithmic terms as 
\be
R^{\Delta}=R^{\Delta^{(0)}}
\left(1+\Delta^{(1)}\log R + \Delta^{(2)} \log R +\frac{1}{2}
{\Delta^{(1)}}^2\log^2 R  + ...\right) \ee 
quite similar to the perturbative logarithms appearing in correlation
functions of local correlators\footnote{We thank T.~Petkou for
a discussion on this point.}. Hence, singling out in the limit
$R \to 0$ the coefficients of the divergent logarithms $\log R$ 
in the connected Green function of two Wilson loops will allow us
to determine the one-loop anomalous dimensions of the operators
appearing in (\ref{OPE}).

Let us briefly comment 
on the relation of the field-theoretic behavior of the expansion (\ref{OPE}) 
with supergravity predictions. 
All local gauge-invariant operators in the ${\cal N}=4$ super Yang-Mills
theory are classified
by irreducible representations of the superconformal group
$SU(2,2|4)$. In particular, operators transforming in short representations,
i.e.~carrying a certain amount of supersymmetry, are
protected from quantum corrections. Conversely, scaling dimensions of long 
superfields are subject to renormalization. 
In the supergravity approximation
the correlator of two loops can be computed by 
evaluating the exchange amplitude of supergravity 
fields between two string worldsheets with  
loops as their boundaries \cite{BCFM}. 
Since supergravity fields are dual to protected operators transforming 
in short representations of the superconformal group,  it is these 
operators that survive in the local operator expansion (\ref{OPE}) 
in the limit of large $N$  and large 't Hooft coupling 
$\lambda$. However a closed Maldacena-Wilson loop
completely breaks supersymmetry.
Therefore its  
operator expansion at finite $\lambda$ should certainly contain
unprotected operators.
We will show that this is indeed the case
and that the operators with non-vanishing anomalous dimensions 
are the only ones coupling to the interacting (i.e.~non-ladder) 
Feynman graphs in the limiting geometry we study.

We start with describing the lower dimensional operator content 
of the expansion (\ref{OPE}) in field theory. 
There are only two types of gauge invariant operators of 
free field dimension two:
the chiral primary operators (CPOs)  $O^a$ 
and the Konishi scalar $K$. 
Canonically normalized
in the large $N$-limit they are given by
\bea
O^a=\frac{4\sqrt{2}
\pi^2}{\lambda}C_{IJ}^a:\Tr(\Phi^I\Phi^J):\, ,
~~~~
K=\frac{4\pi^2}{\sqrt{3}\lambda} :\Tr(\Phi^I\Phi^I):\, .
\eea
Here the traceless symmetric tensor $C_{IJ}^a$ obeys
$C_{IJ}^aC_{IJ}^b=\delta^{ab}$; $a=1,\ldots ,20$. 
The protected operators $O^a$ lie in a short supermultiplet and 
transform in the irrep $\bf{20}$ of the R-symmetry group $SO(6)$. They have vanishing
anomalous dimension. The
Konishi scalar is the lowest component of the long  
supermultiplet \cite{Konishi} and it acquires an anomalous dimension
in perturbation theory \cite{AFGJ, Anselmi}. In particular, its one-loop anomalous
dimension is 
$\Delta^{(1)}_K=\frac{3\lambda}{4\pi^2}$
\cite{Anselmi} which was extensively confirmed in recent studies \cite{BKRS1}-\cite{PS}. 

Among the operators of classical dimension three we have to consider  
\bea 
\label{dim3irrel}
:\Tr(\Phi^I\Phi^J\Phi^J):\, ,~~~~~
:\Tr(\Phi^{\{I}\Phi^J\Phi^{K\}}):\, \eea and \bea
J^I_{\mu\nu}=
:\Tr(\Phi^IF_{\mu\nu}):\, , 
\label{J}
\eea
where $\{,\}$ stands for symmetrization with all traces removed.
In particular the second operator in (\ref{dim3irrel}) is the protected CPO  
transforming in the $\bf{50}$ of $SO(6)$. 
The leading contribution to the correlation function 
of $W$ with any of the operators in (\ref{dim3irrel}) 
is of order  $\lambda^{3}\sim g^6$ and, therefore,  in order to find 
the one-loop anomalous dimensions of these operators
one has to analyze the order $g^8$ contribution
of the correlator $\langle W \, W \rangle$, which is beyond our present knowledge.
As to the operator $J^I_{\mu\nu}$, since the leading order 
behavior is $\langle W J^I_{\mu\nu}\rangle\sim \lambda^2$, 
we are able to use our perturbative results to deduce its anomalous
scaling dimension.

The operator $J^I_{\mu\nu}$ is particularly interesting because its renormalization involves fermions.   
Indeed, the gauged 5-dimensional supergravity contains 
an antisymmetric tensor field $a_{\hat\mu\hat\nu}$
which is dual\footnote{In the formulation of ${\cal N}=4$ SYM 
with Weyl fermions  
$a_{\hat\mu\hat\nu}$ couples to the YM operator $O_{\mu\nu}^+$
of reference \cite{FFZ,AF} transforming in 
the self-dual irrep ${\bf 6}_c$ of $SU(4)$.}
to a dimension three tensor current $J_{\mu\nu}^{I\, +}$ 
belonging to the stress tensor multiplet of the boundary 
conformal field theory \cite{FFZ, AF}. This
 current is not a purely bosonic operator but 
also contains bilinear fermion terms. Hence, although
in the free theory fermions do not couple to the 
Maldacena-Wilson loop, a naive operator $J^I_{\mu\nu}$
does not fit into the local operator expansion at strong coupling. 
Apparently we are encountering  a splitting phenomenon 
which also occurs for operator product expansions of local operators 
\cite{Anselmi,AFP}:  
the free field operator $J^I_{\mu\nu}$ splits
in perturbation theory into the sum of two operators, one of which is a protected
operator whereas the other develops an anomalous scaling dimension and decouples 
at strong coupling.

To justify this picture
define the following two operators  
\bea
J^{I\, \pm}_{\mu\nu}&=&2 \Tr(\Phi^IF_{\mu\nu})\pm\frac{1}{\sqrt{8}}
\Sigma^I_{AB}\Tr(\bar{\psi}^{A} \gamma_{\mu\nu}\psi^{B}) \, .
\eea
Here $\Sigma^I_{AB}$ is an antisymmetric matrix  
which intertwines the adjoint irrep of $SO(4)$ with the fundamental of $SO(6)$, 
$A,B=1,\ldots, 4$.
One can take for instance $\Sigma^I_{AB}=(\eta^k_{AB},\bar{\eta}_{AB}^k)$,
where $k=1,2,3$ and $\eta^k_{AB}$
and $\bar{\eta}_{AB}^k$ are self-dual and anti-self dual 't~Hooft symbols respectively.
In free field theory the operators $J^{I\, \pm}_{\mu\nu}$
are orthogonal with respect to the two-point function 
and satisfy $4J^I_{\mu\nu}=J^{I\,+}_{\mu\nu}+J^{I\,-}_{\mu\nu}$. 
One also has\footnote{The fermionic propagator is 
$
\langle \psi^{a,\, A}_{\alpha}(x) \bar{\psi}^{b,\, B}_{\beta}(y)
   \rangle=-i\delta^{AB}
\frac{\delta^{ab}\gamma_{\alpha\beta}^{\mu}(x-y)_{\mu}}{2\pi^2(x-y)^4}\, .
$}  
\bea
\langle J^{I\, \pm}_{\mu\nu}(x) J^{J\, \pm}_{\lambda\rho}(y)\rangle= 
-\frac{\lambda^2\delta^{IJ}}{\pi^4(x-y)^6}\, \Biggl[\delta_{\mu[\lambda}\delta_{\rho]\nu}   
+2\frac{(x_{\mu}\delta_{\nu[\lambda}-x_{\nu}\delta_{\mu[\lambda})x_{\rho]}}{x^2}
\Biggr]\, . 
\eea
In perturbation theory $J^{I\,+}_{\mu\nu}$ and $J^{I\,-}_{\mu\nu}$ 
have different renormalization group behavior, whereas $J^{I\,+}_{\mu\nu}$
belongs to the stress tensor multiplet and is non-renormalized,
the operator $J^{I\,-}_{\mu\nu}$ has a non-vanishing anomalous scaling 
dimension.

The analysis of the dimension four operators is more involved because they also 
contain descendent fields, which are not orthogonal to primaries with respect to the 
two-point function. 
A given primary operator $O$ contributes to the local operator expansion
with all its derivative descendents. This contribution  
can be found by evaluating the correlation function of $W$ with $O$
\be
\frac{\langle W(C)\,
O(0)\rangle}{\langle W \rangle} =
\sum_{k}c_kR^{\Delta_k} \langle O^{(k)}(x) \,  O(0)\rangle
\ee 
where $k$ enumerates the infinite set of descendents 
of the primary operator $O$.
In particular, the leading contributions of the descendents of the CPO,
of the Konishi scalar and of the currents $J_{\mu\nu}^{\pm}=\theta^IJ_{\mu\nu}^{I\, \pm}$ read as 
\bea
\label{blockCPO}
\frac{\langle W(C)\, O^a (x)\rangle}{\langle W
\rangle}&=&\kappa^a\frac{\lambda\sqrt{2}}{4}
\frac{R^2}{(R^2+h^2)^2}\, ,
\\ 
\label{blockK}
\frac{\langle W(C)\, K(x) \rangle}{\langle W \rangle}&=&
\frac{\lambda}{4\sqrt{3}}\frac{R^2}{(R^2+h^2)^2} \, ,
 \,  \\
\label{blockJ}
\frac{\langle W(C)\,
 J_{\mu\nu}^{\pm}(x) \rangle}{\langle W \rangle}&=&\epsilon_{ij}\delta_{\mu}^i\delta_{\nu}^j\, 
\frac{i\lambda^2}{4\pi^2}\frac{R^3}{(R^2+h^2)^3} \, .
\eea
where $x=(0,0,h,0)$ and the loop of radius $R$ is centered at the origin.
Moreover we have defined $k_a=C_{IJ}^a\theta^I\theta^J$ so that $k_ak^a=5/6$.

Expanding (\ref{blockCPO})-(\ref{blockJ}) in powers of $R^2/h^2$ one can identify every 
monomial as coming from a certain derivative descendent
in the local operator expansion of $W(C)$. 
It is not difficult to find the form of these descendents in the free   
theory. For example to construct the dimension four descendents of the Konishi scalar 
one has to consider the   
following independent operators $:\partial_i\Phi^I\partial_i\Phi^I:$,
$:\partial_{\bar{i}}\Phi^I\partial_{\bar{i}}\Phi^I:$
and  $:\partial_{i}^2\Phi^I\Phi^I:$. There is only one linear combination 
of these operators\footnote{In fact this is a primary operator 
coinciding with the component $T_{ii}$ of the stress tensor $T_{\mu\nu}$ of free six bosons.} 
which is orthogonal to $K$ while the other two provide the descendents we are looking for. 
Since we have several derivative operators of the same 
free field dimension we expect that they will mix under renormalization.
The correlation function of two Maldacena-Wilson loops 
to order $g^6$ does not contain enough information to establish their 
individual one-loop anomalous dimensions. We shall therefore 
restrict the further discussion to the operators of 
approximate dimension two and three mentioned above.  

Summarizing the lower dimensional content of the local operator expansion
we have 
\bea
\label{OPEexact}
\frac{W(C)}{\langle W \rangle}
&=&\uni+R^{\Delta_K}\Biggl[\frac{\lambda}{4\sqrt{3}}+\ldots \Biggr]K(x)+
R^{\Delta_O}\Biggl[\frac{\lambda}{2\sqrt{2}}+\ldots \Biggr]\kappa_a O^a(x) \\
\nonumber
&+&R^{\Delta_+}\Biggl[\frac{i\pi^2}{4}+\ldots\Biggr]J_{ij}^+(x)\epsilon^{ij}
+R^{\Delta_-}\Biggl[\frac{i\pi^2}{4}+\ldots\Biggr]J_{ij}^-(x)\epsilon^{ij}+\mbox{higher}
\, .
\eea
Some comments are in order. In (\ref{OPEexact}) 
the dots indicate higher order terms in $\lambda$
of the corresponding operator expansion coefficients.
Note that the coefficient of the CPO has been computed
in \cite{Semenoff:2001xp} to all orders in $\lambda$ under the assumption
of vanishing radiative corrections.
The  scaling dimensions of the CPO and $J^{+}_{ij}$ are $\Delta_{O}=2$ and $\Delta_{+}=3$
respectively, 
while $\Delta_K$ and $\Delta_-$, being the dimensions of  
the Konishi scalar and of the current $J^{-}_{ij}$,
receive perturbative corrections.
The operator
content of (\ref{OPE}) is sensitive to the orientation of
the loop since the form $\epsilon^{ij}$ flips the sign under change of orientation.
In the following it is convenient to distinguish different
orientations by denoting a contour oriented
clockwise by $C^+$ and the one with the opposite orientation by $C^-$. 

By using eqs.~(\ref{2pt}), (\ref{OPEexact}) and (\ref{blockCPO})-(\ref{blockJ})
we get in the limit $R_1\to 0$ the following leading ${\cal O}(\lambda^2)$ contribution
to the Green function of two Maldacena-Wilson loops: 
\bea 
\nonumber
\frac{\langle W(C_1)W(C_2^{\pm})\rangle_c} {\langle W \rangle^2
}&=&
\frac{\lambda^2}{48}R_1^{\Delta_{K}}\frac{R_2^2}{(R_2^2+h^2)^2} 
+ \frac{5\lambda^2}{48}R_1^{\Delta_{O}}\frac{R_2^2}{(R_2^2+h^2)^2} \nonumber \\
&\mp&
\frac{\lambda^2}{8}R_1^{\Delta_+}\frac{R_2^3}{(R_2^2+h^2)^3} 
\mp\frac{\lambda^2}{8}R_1^{\Delta_-}\frac{R_2^3}{(R_2^2+h^2)^3}
\eea
Here the first line represents the contribution from 
the Konishi scalar and the CPO while the second one is due to $J^{\pm}_{ij}$.
Taking into account the one-loop anomalous dimension 
of the Konishi field we therefore obtain   
\bea 
\label{log}
\frac{\langle W(C_1)W(C^{\pm}_2)\rangle_c} {\langle W \rangle^2
}&=&\frac{\lambda^2}{8}\frac{R_1^2R_2^2}{(R_2^2+h^2)^2}
\mp\frac{\lambda^2}{4}\frac{R_1^3R_2^3}{(R_2^2+h^2)^3} \\
&+&\frac{\lambda^3}{64\pi^2}\frac{R_1^2R_2^2}{(R_2^2+h^2)^2} \log R_1 
\mp 
\frac{\lambda^3}{8}\Delta_{-}^{(1)}\frac{R_1^3R_2^3}{(R_2^2+h^2)^3} \log R_1
 \, .
\nonumber
\eea
This formula predicts that the correlator (\ref{2loopcor}) develops 
a logarithmic singularity in the limit $R_1\to 0$ at order $\lambda^3$. 

Now we are ready to compare the expected behavior of eq.~\eqn{log},
with the explicit two-loop calculation of \cite{Plefka:2001bu}. 
We first note that in the limit considered,
$R_1\to 0$, only the interacting diagrams give rise
to logarithmic terms, on which we shall focus. 
In agreement with the above predictions the contribution
of the graphs can be worked out explicitly from the corresponding
integral representations in this limit. We find
for two loops of equal orientation the logarithmic contributions
\bea
\Biggl [\same{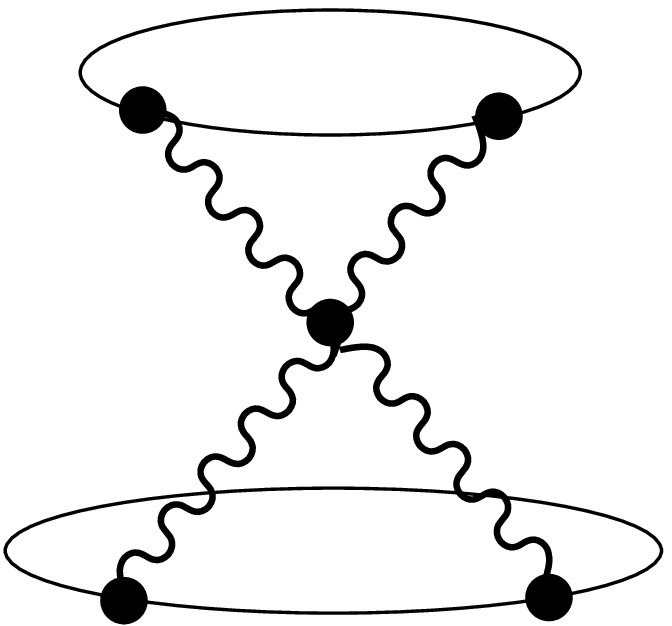}\Biggr]_{R_1\to 0}&=& \frac{\lambda^3}{64\pi^2}\,
  \frac{R_1^3R_2^3}{(R_2^2+h^2)^3}   \log R_1\nn \, ,\\
&&\nn\\
\Biggl [ \same{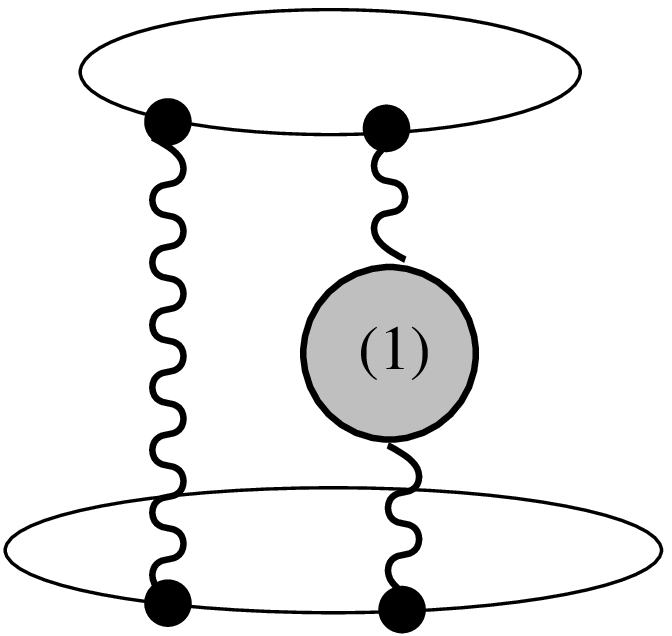} + \same{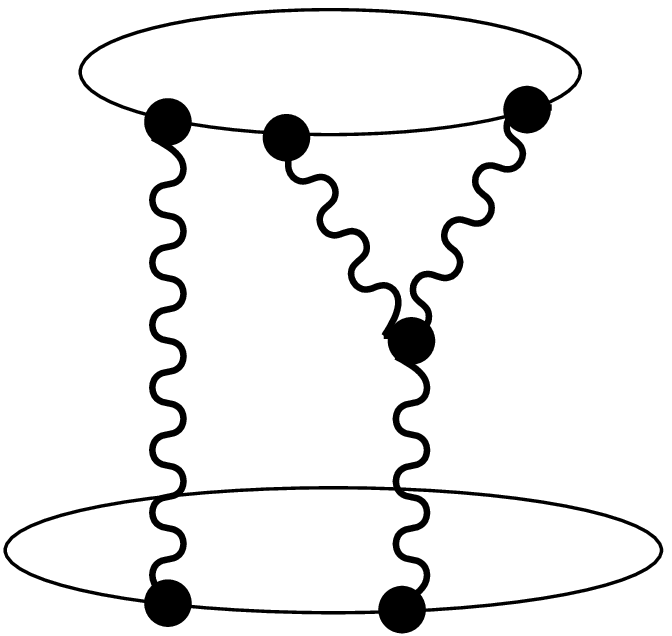}\phantom{\Biggr]_{R_1\to 0}}&=& 
\frac{\lambda^3}{32\pi^2}\, 
\frac{R_1^2R_2^2}{(R_2^2+h^2)^2}\log R_1 -
\frac{7\,\lambda^3}{64\pi^2}\frac{R_1^3R_2^3}{(R_2^2+h^2)^3}\,
\log R_1 \nn \\ +\same{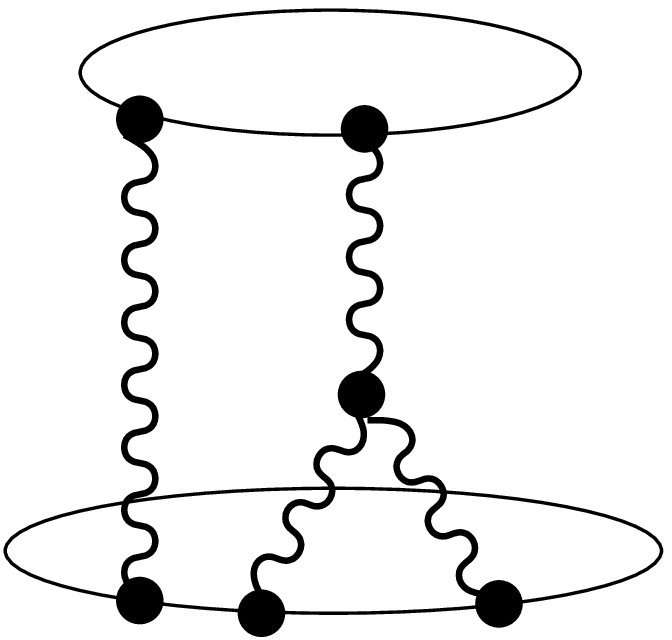} \Biggr]_{R_1\to 0}
&-&\frac{11\lambda^3}{96\pi^2}\frac{R_1^4R_2^2}{(R_2^2+h^2)^3}\log R_1 \, ,
\nonumber
\label{OPEgraphs}\\
&&\nn\\
\Biggl [ \same{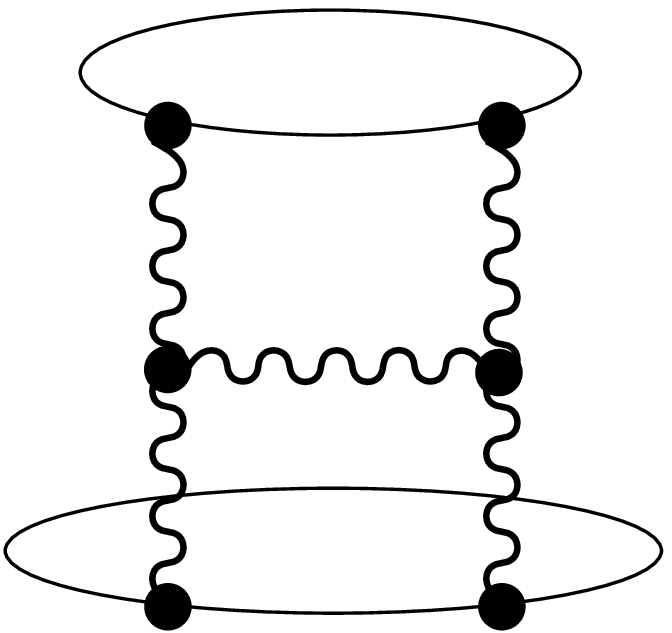} \Biggr ]_{R_1\to 0}&=& 
-\frac{\lambda^3}{64\pi^2}\, 
\frac{R_1^2R_2^2}{(R_2^2+h^2)^2}\log R_1 +
\frac{\lambda^3}{32\pi^2}\,
\frac{R_1^3R_2^3}{(R_2^2+h^2)^3}\log R_1 \nn \,  
\nonumber \\
&+&\frac{59\lambda^3}{1152\pi^2}\frac{R_1^4R_2^2}{(R_2^2+h^2)^3}\log R_1 \, .
\eea
The ladder graphs do not yield logarithmic contributions.
Summing up we obtain 
\bea
\nonumber
\frac{\langle W(C)W(C)\rangle_c} {\langle W \rangle^2
}&=&
\frac{\lambda^3}{64\pi^2}\frac{R_1^2R_2^2}{(R_2^2+h^2)^2} \log R_1 
-\frac{\lambda^3}{16\pi^2}\frac{R_1^3R_2^3}{(R_2^2+h^2)^3} \log R_1 \\
&-&\frac{73\lambda^3}{1152\pi^2}\frac{R_1^4R_2^2}{(R_2^2+h^2)^3}\log R_1
\, .
\label{logdiag}
\eea
The result for graphs of opposite orientation differs from
\eqn{OPEgraphs} only by a sign in front of the 
$({R_1R_2})^3\log R_1$ terms.
The first term in the last expression 
indeed confirms the one-loop anomalous
dimension of the Konishi field to be $\Delta_K^{(1)}=\frac{3\lambda}{4\pi^2}$.
By comparing the second term in (\ref{logdiag}) 
with the local operator expansion predictions (\ref{log})
we read off the one-loop anomalous dimension of the operator $J^{I\, -}_{ij}$ 
\be
\Delta_{-}^{(1)}=\frac{\lambda}{2\pi^2}\, .
\ee 
Finally the last term in (\ref{logdiag}) carries information
on the  one-loop anomalous dimensions of the approximate dimension four 
operators, which are not orthogonal to the Konishi scalar and the CPO. 

This completes our considerations of the perturbative local operator 
expansion. We conclude by emphasizing two observations 
concerning the above analysis. 
First, the interacting Feynman diagrams are the ones 
responsible for the appearance of unprotected operators in the 
local operator expansion. 
The ladder diagrams
do not produce any logarithmic singularities (at the level of the perturbation 
theory we are working at) and contribute only to the local operator
expansion coefficients.
In this respect it would be interesting to understand whether the interacting
graphs also contribute to the local operator expansion coefficient of the CPOs. 
If this is not the case this would mean that interacting graphs couple solely 
to the unprotected operators of the theory, like the Konishi field. 
At strong coupling the Konishi field 
decouples from the theory, which would imply in turn that the 
sector of interacting graphs coupling to it
becomes negligible in comparison to the ladders\footnote{On the other hand,
the sector of interacting graphs coupling to unprotected operators 
of {\it finite} anomalous dimension at strong
coupling will survive.}.  
Unfortunately the information we have at our disposal so far does not 
suffice to answer this question.
Second, in order to match the supergravity picture the operator 
expansion of the bosonic Maldacena-Wilson loop 
should contain renormalized operators which are constructed from  
bosons and fermions. Fermions emerge due to the splitting mechanism and 
at the level of perturbation theory
this can be viewed as a sign 
of supersymmetry in the theory of a purely bosonic operator.

\section{New Results on the Matrix Model Conjecture}

Armed with the complete order $g^6$ result for the connected
correlator of two Maldacena-Wilson loops of equal orientation,
we may now perform a further test of the Drukker-Gross
matrix model conjecture \cite{Drukker:2000rr}.  
Based on an anomaly argument these authors argue that
the expectation value of an arbitrary number of
coincident, possibly multiply wound,
circular Maldacena-Wilson loops of equal orientation
is given by solving a purely combinatorial problem\footnote{
This more general problem is equivalent to finding the
expectation value of a single circular Wilson loop in an 
arbitrary representation of the gauge group.}.  
The latter is easily seen to correspond to computing 
correlators in a simple Gaussian matrix model.
If correct this conjecture amounts to the claim that
all interacting graphs contributing to the loop correlator
vanish.

The relevant limit in order to extract the result of two incident 
circles of equal orientation from our geometry is
\be
R_1=R_2=R \qquad \mbox{and} \qquad h\rightarrow 0 \, .
\label{hgleich0}
\ee
Given the explicit integral representations for individual diagrams
found in \cite{Plefka:2001bu} it is straightforward to analytically
perform this limit.
Remarkably, and quite non-trivially, one finds after some calculation
that {\it all} interacting non-ladder graphs cancel:
\bea
0&=&\Bigg [\, \same{SE}+\same{IY}+\same{IYT}\, \Bigg ]_{\stackrel{R_1=R_2}
{h\rightarrow 0}}
\nn\\
0&=&\Bigg [\, \same{X}\, \Bigg ]_{\stackrel{R_1=R_2}{ h\rightarrow 0}} \nn\\
0&=&\Bigg [\, \same{H}\, \Bigg ]_{\stackrel{R_1=R_2}{ h\rightarrow 0}}
\label{SameInt}
\eea
Therefore only the contribution of the ladder-diagrams survives.
This result represents a two loop order test of the 
Drukker-Gross conjecture \cite{Drukker:2000rr}.
For the ladder-graphs  one has in the limit \eqn{hgleich0}
\bea
\frac{\lambda}{4} &=&
\Bigg [\, \same{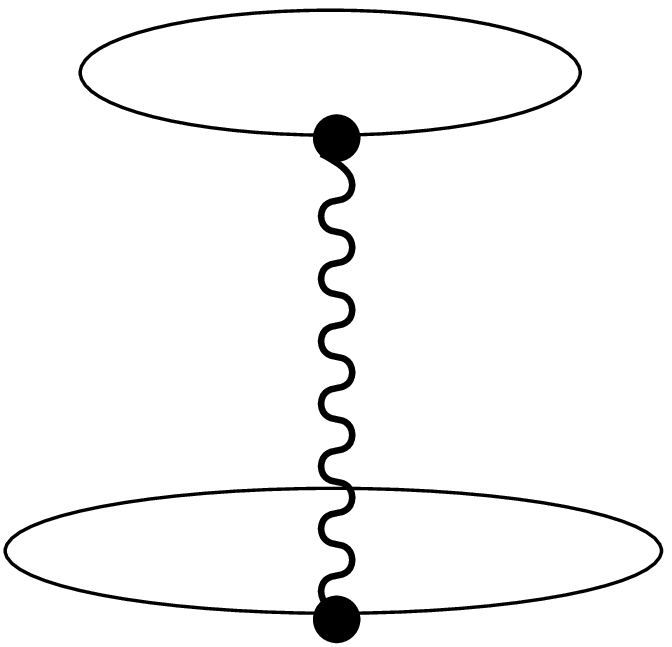}\, \Bigg ]_{\stackrel{R_1=R_2}{h\rightarrow 0}}
 \nn\\
\frac{3\, \lambda^2}{32} &=&
 \Bigg [\,\same{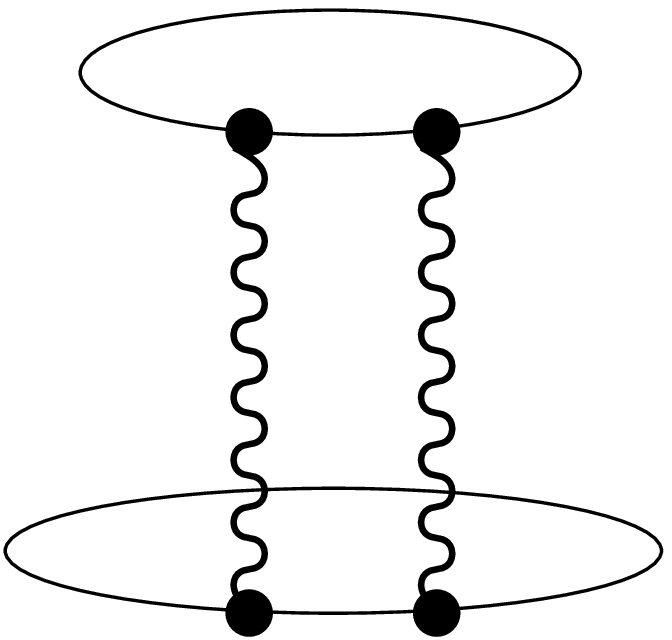} + \same{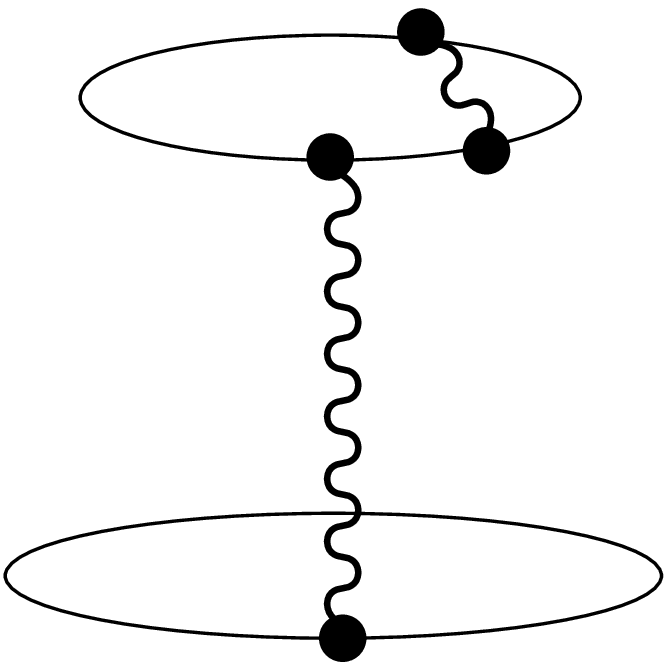} + \same{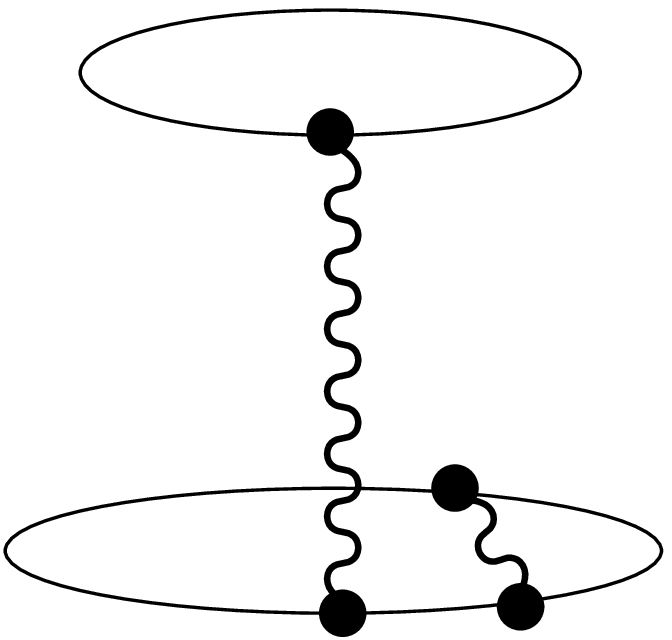}
\, \Bigg ]_{\stackrel{R_1=R_2}{h\rightarrow 0}}
\nn\\
\frac{5\, \lambda^3}{384} &=& 
\Bigg [\,\same{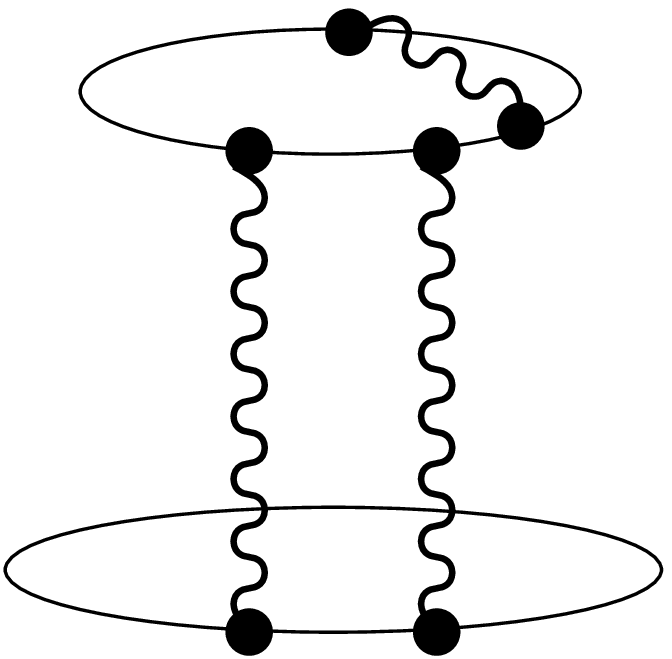}+  \same{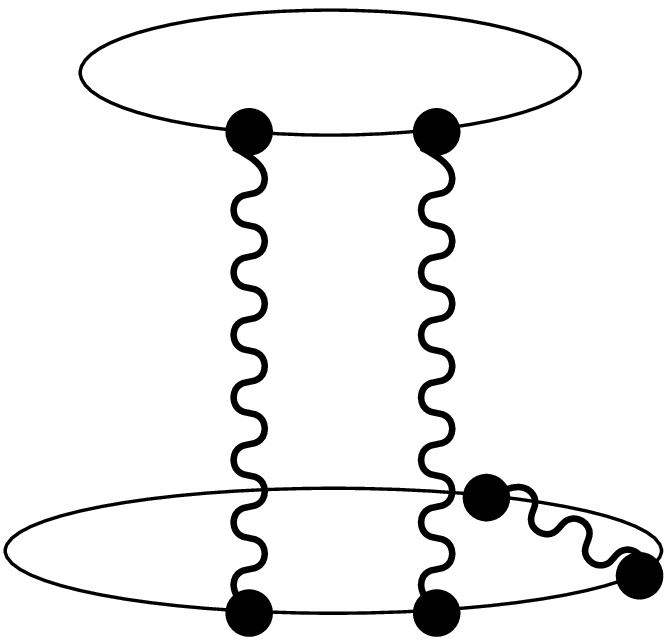}+
\same{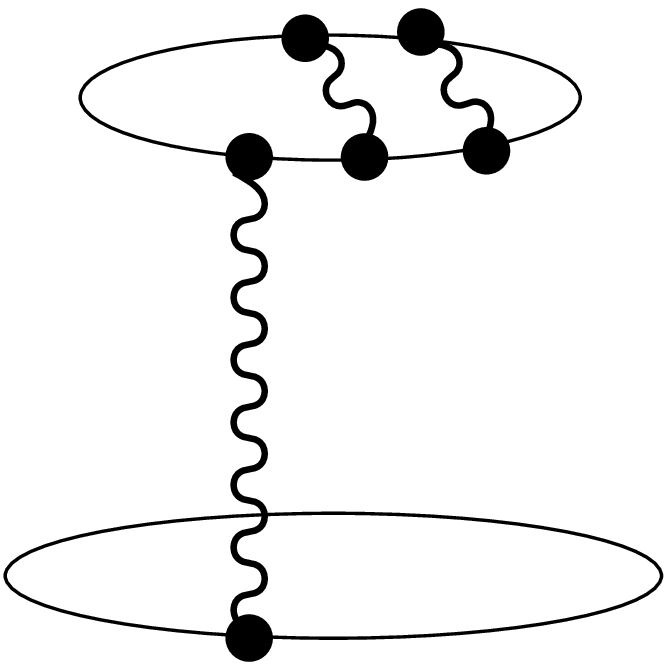} +\same{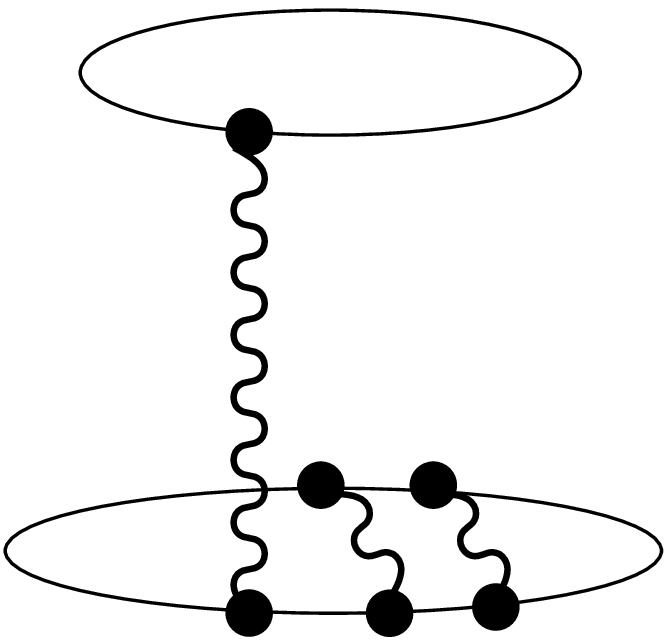} + \same{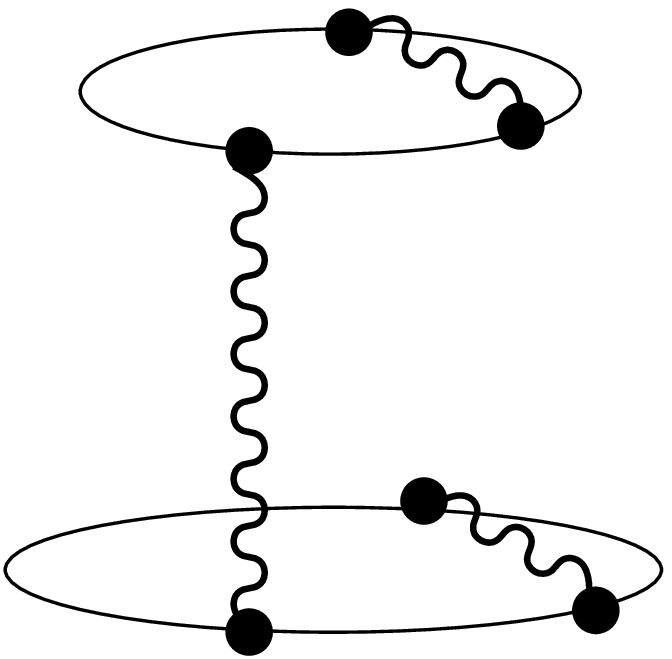}
+\same{3Gluon}\, \Bigg ]_{\stackrel{R_1=R_2}{h\rightarrow 0}}
  \nn\\
 \frac{\lambda^3}{192\, N^2} &=&
\Bigg [\, \same{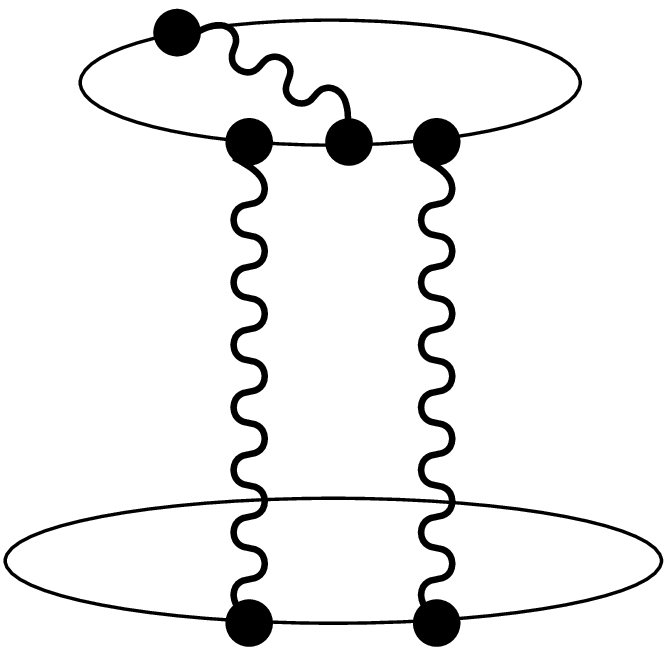} +\same{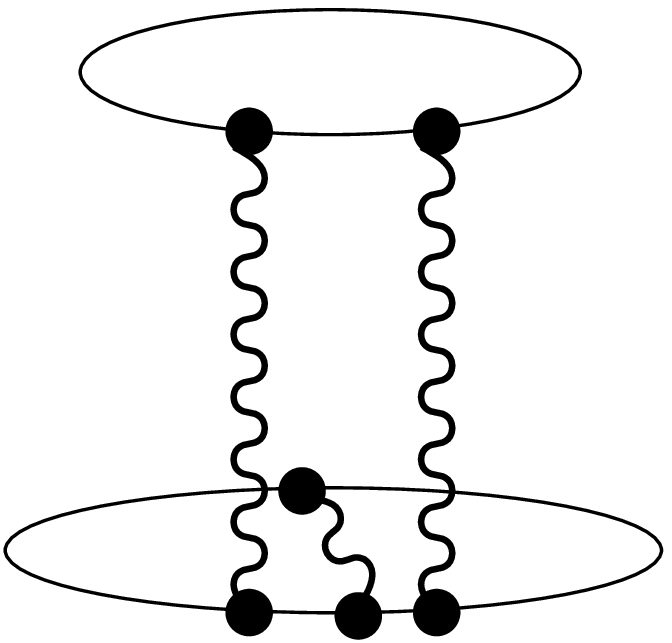} +\same{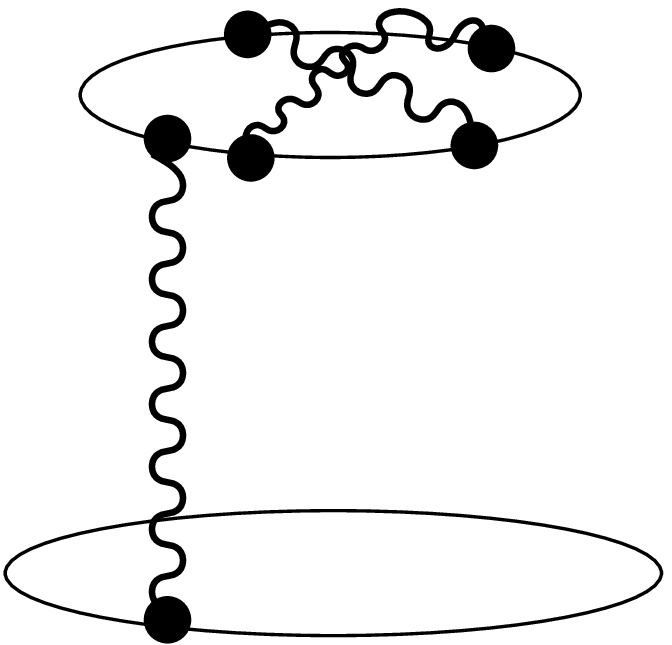}+
\same{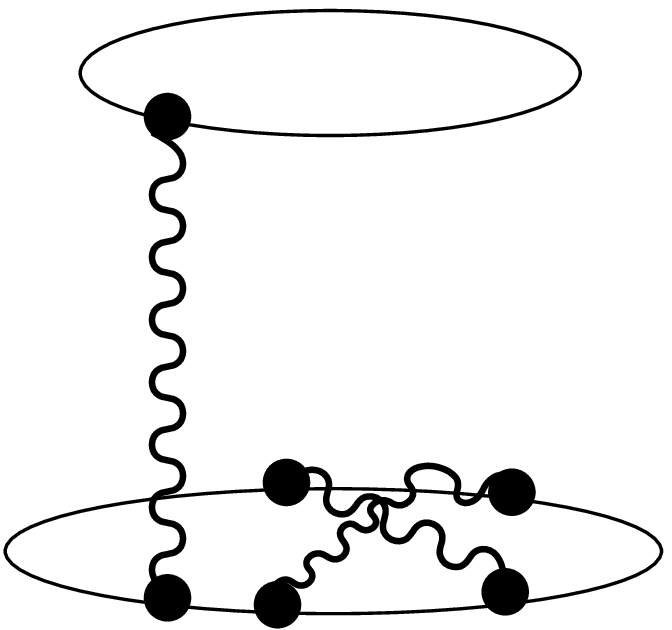}+ \same{3Gluon-NP}
\, \Bigg ]_{\stackrel{R_1=R_2}{h\rightarrow 0}}
\eea
which indeed follows from the proposed matrix model description 
\cite{Akemann:2001st,gernot}
\bea
\langle \W(C_1)\, \W(C_2) \rangle_{c} &=&
\frac{1}{Z}\int dM\, \Tr e^M \, \Tr e^M\, e^{-\frac{2\, N}{\lambda}\, \Tr M^2}
-\Bigl (\frac{1}{Z}\int dM\, \Tr e^M \, e^{-\frac{2\, N}{\lambda}\, \Tr M^2}
\Bigr )^2 \nn\\
&=& \frac{\sqrt{\lambda}}{2}\,I_0(\sqrt{\lambda})I_1(\sqrt{\lambda})
+ \frac{\lambda^2}{192\, N^2}\,
\Bigl( 3 I_1(\sqrt{\lambda})^2 \nn\\&& \qquad
+ I_2(\sqrt{\lambda}) [ 2 I_0(\sqrt{\lambda})
+ I_2(\sqrt{\lambda})\, ] \, \Bigr )
+ {\cal O}(\frac{1}{N^4})
\label{MModel}
\eea
where $I_n(x)$ is the $n$th order modified Bessel function.
Expanding \eqn{MModel} reproduces our findings 
to ${\cal O}(\lambda^3)$.

Let us review the existing perturbative evidence for 
the matrix model conjecture.
Clearly, performing the sum of all ladder graphs for correlators
of incident Maldacena-Wilson circles may be rephrased by
a Gaussian matrix model of the form \eqn{MModel}. This is
due to the fact that the combined gluon-scalar propagator
trivializes to a constant if both of its ends lie on a
circle. An alternative way of understanding this starts
from the observation that the ladder graphs of an
infinite Maldacena-Wilson line are identically zero.
Under an inversion the line is mapped to a circle,
so one might expect the correlators of the line
and the circle to be identical by conformal invariance
of the theory.
However, as we know, this is not
the case, and as shown in \cite{Drukker:2000rr} 
the discrepancy is
due to additional total derivative terms
that the gluon propagator picks up under an inversion.
These total derivatives let the contribution
from the {\it ladder} diagrams collapse to a point 
and turn the circle correlator into a correlator
in a zero dimensional field theory, the matrix model
\eqn{MModel}.
The story for the interacting graphs is, however,
considerably more intricate\footnote{Here we cannot follow the
arguments of \cite{Drukker:2000rr} which claim that
also the interacting sector collapses to a point.
Even stronger, the authors speculate that there are
no contributions from the interacting graphs to
all orders in perturbation theory.}. Now at least one leg of 
the modified propagator
is attached to an interaction vertex which may no longer be
simply integrated by parts. 
An explicit example where
such a graph by graph consideration fails
appears already at the one loop (${\cal O}(g^4)$) 
level \cite{Erickson:2000af}. 
Here the interacting graphs for the 
infinite Maldacena-Wilson line vanish separately
\be
\raisebox{-.2cm}{\epsfxsize=2.4cm\epsfbox{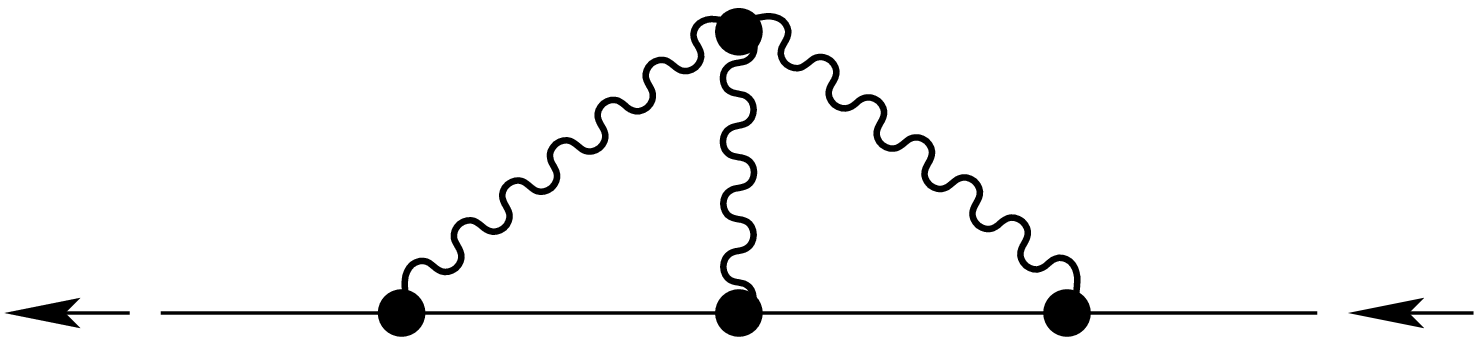}}\,
=0 \qquad
\raisebox{-.2cm}{\epsfxsize=2.4cm\epsfbox{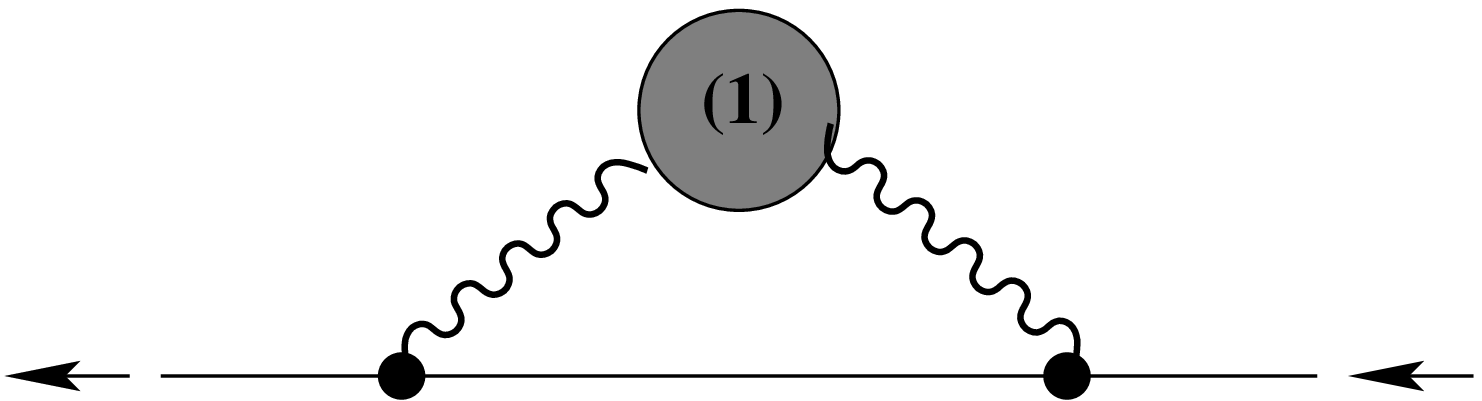}}\,
=0
\ee
due to the appearance of the standard combination
$\dot x_i\cdot \dot x_j - |\dot x_i|\,|\dot x_j|$
in the numerators, which vanishes for straight
line trajectories. Their image graphs under inversion,
however, do not vanish and are actually individually
divergent in four dimensions \cite{Erickson:2000af}
\be
\raisebox{-0.5cm}{\epsfxsize=1.2cm\epsfbox{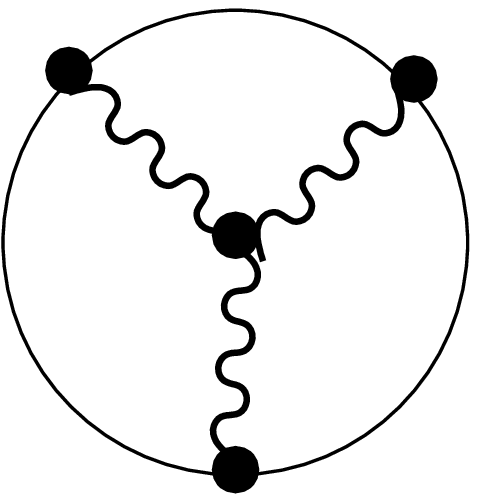}}
= \mbox{divergent} \qquad
\raisebox{-0.5cm}{\epsfxsize=1.2cm\epsfbox{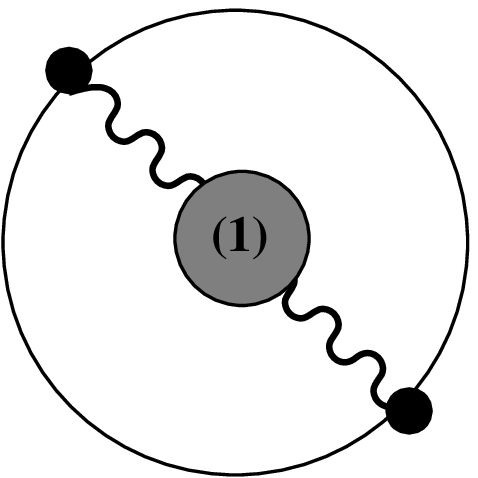}}
= \mbox{divergent} 
\label{radoneloop}
\ee
It is only that the sum of these two graphs vanishes,
due to the subtle cancellation of bulk (right diagram)
versus boundary (left diagram)
divergences for the Maldacena-Wilson loop operator.
Given this, one might speculate on the existence
of a modified anomaly argument, in which one only 
considers the inversion of classes of individually
divergent graphs, which sum to a finite result. This is indeed a necessity
in view of the fact that the dimensional 
regularization needed for divergent graphs breaks 
conformal invariance. If true and if the resulting
matrix model is indeed Gaussian, such a modified conjecture
would claim that the image graph under inversion of a 
vanishing interacting graph on the line should either
be zero or divergent. Indeed the results of \cite{Erickson:2000af}
in \eqn{radoneloop} and our reported two-loop findings of
\eqn{SameInt} support this statement.

However, let us now give a counterexample to such a
modified anomaly conjecture. Consider the four point
insertion on the line and the circle, an order $g^6$
interacting graph. On the circle it takes the
form
\bea
\raisebox{-0.5cm}{\epsfxsize=1.2cm\epsfbox{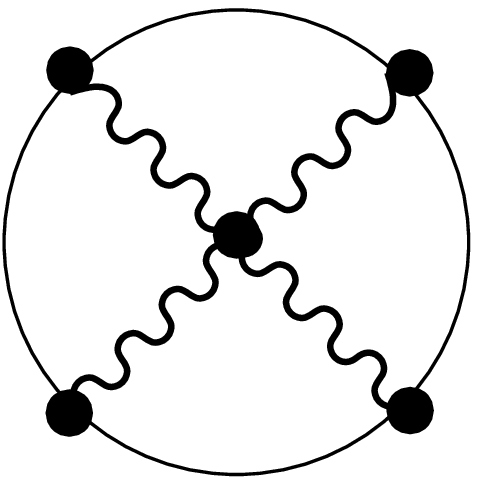}}&=&
\frac{g^6N^2(N-1)}{8}\int \frac{d^4z}{(2\pi)^8}
\int_{0}^{2\pi}d\tau_1\int_{0}^{\tau_1}d\tau_2\int_o^{\tau_2}d\tau_3
\int_0^{\tau_3}d\tau_4 \times \nn\\
&\times& \frac{2\cdot(2,4)\cdot(3,1)-(1,4)\cdot(2,3)-(1,2)\cdot(3,4)}
{(x_1-z)^2\, (x_2-z)^2\, (x_3-z)^2\,
(x_4-z)^2} 
\label{rad1}
\eea
where we have parametrized the four points on the circle by
$x_i=(\cos\tau_i,\sin\tau_i,0,0)$ and introduced the notation
$(i,j):=\xd{i}\cdot\xd{j}-|\xd{i}|\,|\xd{j}|$. From this
structure one immediately deduces the vanishing of the
corresponding graph on the line
\be
\raisebox{-.2cm}{\epsfxsize=2.4cm\epsfbox{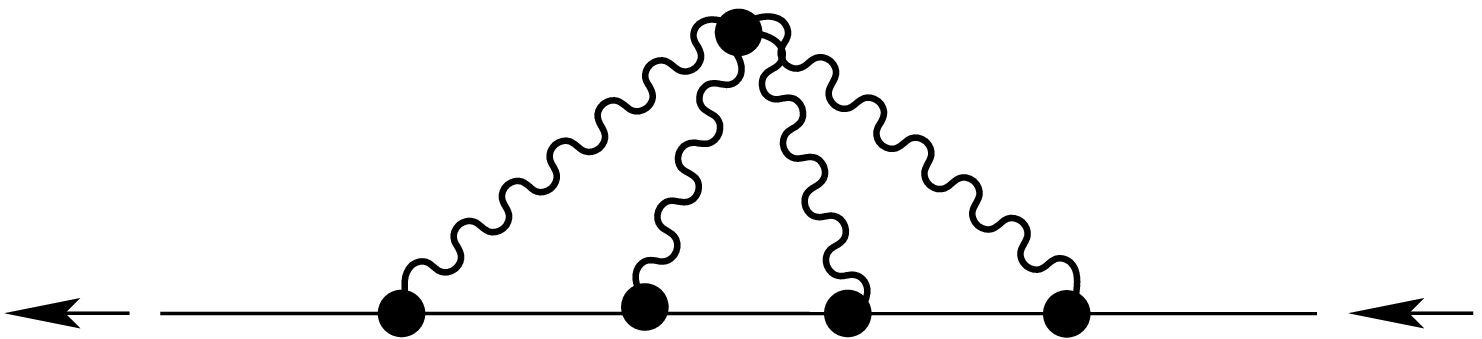}}\,
=0 
\ee
due to $(i,j)|_{\mbox{\footnotesize line}}=0$. The circle
graph \eqn{rad1} on the other hand may be brought into
the form
\bea
\raisebox{-0.5cm}{\epsfxsize=1.2cm\epsfbox{Rad.eps}}&=&
\frac{g^6N^2(N-1)}{8\,(2\pi)^6}\int_0^\infty dz\int_0^\infty d\rho
\int_{0}^{2\pi}d\tau_1\int_{0}^{\tau_1}d\tau_2\int_o^{\tau_2}d\tau_3
\int_0^{\tau_3}d\tau_4  \times \nn\\
&&\times \rho\,z\,\frac{2\cdot[2,4]\cdot[3,1]-[1,4]\cdot[2,3]-[1,2]\cdot[3,4]}
{\prod_{i=1}^4(1+\rho^2+z^2-2z\cos\tau_i)}
\eea
where $[i,j]:=(i,j)|_{\mbox{\footnotesize circle}}=1-\cos(\tau_i-\tau_j)$. 
We claim that this integral is
finite. Unfortunately we have not yet been able to determine its
value analytically, but could evaluate it to a high degree of
accuracy numerically. We find that
\be
\raisebox{-0.5cm}{\epsfxsize=1.2cm\epsfbox{Rad.eps}}=\frac{g^6N^2(N-1)}{2^7}
\,\cdot \,9.25(47)\,\cdot 10^{-3}\, \simeq \, \frac{g^6N^2(N-1)}{(4!)^3} 
\ee
where the last value represents our sophisticated guess for the
exact value of the integral.

This consideration shows that the structure of the interacting graphs
at order $g^6$ for a single Maldacena-Wilson loop has a much
richer structure than one would expect from the $g^4$ results.
Although we have only considered a single graph, its finiteness 
casts some doubt on the conjecture \cite{Drukker:2000rr}
that the circular loop can be obtained
by simply solving a combinatorial problem, described by a
Gaussian matrix model. Since it is far from clear how to 
obtain the above X-graph by pure combinatorial reasoning 
it is, furthermore, not obvious whether introducing interactions
into the matrix model might help. 
Of course an individual graph such as the one we just
studied is not gauge invariant by itself, but that is also true
for the artificial - in the context of non-abelian gauge
theory - restriction to ladder graphs. 
Nevertheless, it might still be true that the {\it sum} of all
vertex graphs, in Feynman gauge, cancels. If so, we feel we have demonstrated
that the true mechanism is wanting. 
Clearly it would be highly desirable to
know the complete ${\cal O}(g^6)$ result and to really
understand the behavior of the interacting graphs under
the inversion.

\section*{Acknowledgments.}
It is a pleasure to thank Gernot Akemann,
Nadav Drukker, Tassos Petkou, Gordon Semenoff and Stefan Theisen
for interesting discussions. G.~A. was supported by the DFG and by
the European Commission RTN programme HPRN-CT-2000-00131 in which he is
associated to U. Bonn, and in part by RFBI grant N99-01-00166 and by
INTAS-99-1782.

\end{document}